\documentclass[aps,pra,twocolumn,showpacs,superscriptaddress]{revtex4-1}
\usepackage{amsmath,amssymb}
\usepackage{graphicx}
\usepackage{natbib}
\usepackage{bm}

\begin{document}
\title{Doublon production rate in modulated optical lattices}

\author{Akiyuki Tokuno}
\affiliation{DPMC-MaNEP, University of Geneva, 24 Quai Ernest-Ansermet
CH-1211 Geneva, Switzerland.}

\author{Eugene Demler}
\affiliation{Department of Physics, Harvard University, Cambridge, MA
02138.}

\author{Thierry Giamarchi}
\affiliation{DPMC-MaNEP, University of Geneva, 24 Quai Ernest-Ansermet
CH-1211 Geneva, Switzerland.}

\date{\today}

\begin{abstract}
 We study theoretically lattice modulation experiments with ultracold
 fermions in optical lattices. 
 We focus on the regime relevant to current experiments when interaction
 strength is larger than the bandwidth and temperature is higher than
 magnetic superexchange energy. 
 We obtain analytical expressions for the rate of doublon production as
 a function of modulation frequency, filling factor, and temperature.
 We use local density approximation to average over inhomogeneous
 density for atoms in a parabolic trap and find excellent agreement with
 experimentally measured values. 
 Our results suggest that lattice modulation experiments can be used for
 thermometry of strongly interacting fermionic ensembles in optical
 lattices.
\end{abstract}


\maketitle

\section{Introduction}

Cold atoms provide a new platform in which one can explore long standing
open questions of strongly correlated systems in condensed matter
physics~\cite{Bloch.Dalibard.Zwerger/RevModPhys80.2008:_review,Esslinger/AnuuRevCondMattPhys1.2010:review}.
In particular, two-component Fermi mixtures in an optical lattice provide
an ideal realization of the fermionic Hubbard model, where two species
of fermions -- corresponding to the spin $1/2$ -- interact with an
on-site repulsion.
This model is relevant for understanding properties of electrons in
several classes of novel quantum materials including oxides and
high-$T_c$ superconductors~\cite{Imada.Fujimori.Tokura/RevModPhys70.1998:review,Lee.Nagaosa.Wen/RevModPhys78.2006:review}.
There are currently many efforts to probe the low-temperature physics
of such a model, with experiments already demonstrating the Mott
insulating behavior expected in this
model~\cite{Joerdens.etal/nature455.2008,Schneider.etal/science322.2008}.

An important feature of strongly correlated ultracold atoms is
that traditional probes used in solid state physics are often not
readily available.  
One thus needs to understand how experimental techniques appropriate
for atomic ensembles can provide information on many-body states. 
In this paper we focus on understanding lattice modulation experiments
with fermions in optical lattices. 
The technique of lattice modulation was originally introduced for
bosonic systems and absorbed energy was measured as a function of
modulation frequency~\cite{Stoeferle.etal/PRL.92.2004}.
Measuring energy absorption, however, can not be done accurately enough
for strongly interacting fermions.
Thus an extension of this technique was
proposed,~\cite{Kollath.etal/RPA74R.2006} and 
implemented~\cite{Greif.etal/PRL.106.2011}, in which the number of
doubly occupied sites created by the lattice modulation was measured.
Recent experiments successfully reached the regime of weak perturbations
in which the number of doublons created scales quadratically with the
modulation amplitude (Fermi's golden rule) and the modulation pulse
duration~\cite{Greif.etal/PRL.106.2011}.

While theoretical understanding of such experiments with bosons is
now relatively
complete~\cite{Reischl.etal/PRA72.2005,Iucci.etal/PRA73R.2006,Kollath.etal/PRL97.2006},
the case of fermions turned out to be more challenging. 
The main difficulty is the presence of excitations at very different
energy scales: high-energy charge excitations, so-called doublons and
holons, that have energies set by the on-site repulsion $U$ and fermion
hopping strength $J$, and magnetic excitations that have energies of the
order of superexchange energy $J^2/U$. 
Understanding the interplay of charge and spin degrees of freedom in the
Hubbard model is a long standing problem in condensed matter
physics~\cite{Brinkman.Rice/PRB2.1970,Kane.Lee.Read/PRB39.1989}.
In the special case of half filling and fully disordered spin states,
analysis of lattice modulation experiments has been performed
previously~\cite{Sensarma.etal/PRL103.2009,Huber.Ruegg/PRL.102.2009}. 
However, such analysis is not sufficient for quantitative comparison to
experiments which are done with systems in a parabolic potential that
have a large number of atoms outside of the incompressible Mott
plateaus.

Furthermore such real-time dynamics at finite temperature as seen in
current experiments is problematic for numerics.
Monte Carlo simulations~\cite{Xu.et.al/PRA84R.2011} suffer from the
problems of analytic continuation.
Density-matrix renormalization group approaches, which can deal with
real time dynamics, are so far limited to one dimension.
Therefore, a comparison of dynamical quantities to experiments is a
non-trivial theoretical subject. 

In this paper we develop another analytical approach to calculate
finite-temperature dynamics, that is, the doublon production rate (DPR),
based on the slave-particle 
technique~\cite{Kotliar.Ruckenstein/PRL57.1989,Sensarma.etal/PRL103.2009,Ruegg.Huber.Sigrist/PRB.81.2010}. 
This approach is particularly adapted to the paramagnetic phase of the
Hubbard model, and can be applied to any filling of the band and finite
temperatures whose region is relevant to ongoing experiments. 
It provides a remarkable agreement to the experiments and allows for
potential extensions.

This paper is organized as follows. 
We define the system Hamiltonian, and introduce the slave-particle
representation in Sec.~\ref{sec:slave-part-repr}. 
In Sec.~\ref{sec:diagrammatic-analys}, the spectral functions of the
slave particles are evaluated, and in addition the spectral function of
the original fermionic atom is also obtained.
In Sec.~\ref{sec:DPR}, we proceed with the estimation of the doublon
production rate based on the obtained spectral functions in
Sec.~\ref{sec:diagrammatic-analys}, and
the analytic formula of the DPR spectrum is given.
In Sec.~\ref{sec:comparison-experiments}, the obtained analytic formula
is extended to an inhomogeneous system in a trap by the local density
approximation (LDA), and a comparison with the experiment in
Ref.~\cite{Greif.etal/PRL.106.2011} is implemented. 
Finally, Sec.~\ref{sec:summary} is devoted to the summary.

\section{Slave particle representation}\label{sec:slave-part-repr}
We consider the Hubbard model $H_0=H_{\rm K}+H_{\rm at}$ with
\begin{subequations}
 \begin{align}
  & H_{\rm K}
  =-J\sum_{\sigma,\langle{i,j}\rangle}c^{\dagger}_{i\sigma}c_{j\sigma},
  \\
  & H_{\rm at}
  =-\mu\sum_{j,\sigma}n_{j\sigma}+U\sum_{j}n_{j\uparrow}n_{j\downarrow},
  \label{eq:atomic-Hamiltonian}
 \end{align}
 \label{eq:Hubbard_model}
\end{subequations}
where $c^{\dagger}_{j\sigma}$ and
$n_{j\sigma}=c^{\dagger}_{j\sigma}c_{j\sigma}$ are, respectively, a
creation and number operator of a spin-$\sigma$ fermions at a $j$th site.
We use the following slave-particle
representation: 
\begin{equation}
 c^{\dagger}_{j\uparrow}
 =b^{\dagger}_{j\uparrow}h_{j}+b_{j\downarrow}d^{\dagger}_{j},
 \qquad
 c^{\dagger}_{j\downarrow}
 =b^{\dagger}_{j\downarrow}h_{j}-b_{j\uparrow}d^{\dagger}_{j},
 \label{eq:SF-SB}
\end{equation}
where $b^{\dagger}_{j\sigma}$, $h^{\dagger}_{j}$ and $d^{\dagger}_{j}$
are, respectively, creation operators of a slave boson of
spin-$\sigma$ state (Schwinger boson), holon and doublon at a site $j$.
They satisfy the (anti-)commutation relations,
$[b_{i\sigma},b^{\dagger}_{j\sigma'}]=\delta_{i,j}\delta_{\sigma,\sigma'}$
and $\{h_i,h_j^{\dagger}\}=\{d_i,d_j^{\dagger}\}=\delta_{i,j}$.
The enlarged Hilbert space is projected onto the physical one by
the following constraint at every site: 
\begin{equation}
 \sum_{\sigma}b^{\dagger}_{j\sigma}b_{j\sigma}
 +h^{\dagger}_{j}h_{j} 
 +d^{\dagger}_{j}d_{j} 
 =1.
 \label{eq:Constraint}
\end{equation}
Equation~(\ref{eq:SF-SB}) allows us to rewrite the
Hamiltonian~(\ref{eq:Hubbard_model}) as
\begin{align}
 &H_{\rm K}
   =J\sum_{\langle i,j\rangle}
        \left[
          F_{ji}(h_{i}^{\dagger}h_{j}-d_{i}^{\dagger}d_{j})
          +(A_{ij}^{\dagger}d_{j}h_{i}+{\rm H.c.})
        \right],
 \label{eq:TunnelingHamiltonian} \\
 &H_{\rm at}
   =\sum_{j}
      \left[
        \epsilon^{\rm d}_{j}d^{\dagger}_{j}d_{j}
        +\epsilon^{\rm h}_{j} h^{\dagger}_{j}h_{j}
        +\sum_{\sigma}\epsilon^{\rm b}_{j} b^{\dagger}_{j\sigma}b_{j\sigma}
        -\lambda_j
      \right],
\label{eq:AtomicHamiltonian}
\end{align}
where
$F_{ji}=\sum_{\sigma}b^{\dagger}_{j\sigma}b_{i\sigma}$ and
$A^{\dagger}_{ij}=b^{\dagger}_{i\uparrow}b^{\dagger}_{j\downarrow}-b^{\dagger}_{j\downarrow}b^{\dagger}_{i\uparrow}$
mean the hopping of slave bosons and the creation of
spin singlet pair, respectively.
The local potentials of a slave boson, holon, and doublon are,
respectively, defined as
$\epsilon^{\rm b}_{j}=\lambda_j$,
$\epsilon^{\rm h}_{j}=\mu+\lambda_j$, and
$\epsilon^{\rm d}_{j}=U-\mu+\lambda_j$.
The constraint~(\ref{eq:Constraint}) is implemented via the Lagrange
multiplier $\lambda_j$.

\section{Diagrammatic analysis}\label{sec:diagrammatic-analys}
\subsection{Atomic limit}
We start with the atomic limit ($J/U=0$).
Then the kinetic term~(\ref{eq:TunnelingHamiltonian}) which describes
the scattering among slave particles vanishes.
Since the atomic Hamiltonian~(\ref{eq:AtomicHamiltonian}) is quadratic,
the atomic propagators at $j$th site are easily obtained as
\begin{equation}
 {\cal G}^{(0)}_{{\rm b}\sigma}(\bm{r}_j,i\omega_n)
 =\frac{1}{i\omega_n-\epsilon^{\rm b}_{j}},
 \quad
 {\cal G}^{(0)}_{\rm d/h}(\bm{r}_j,i\nu_n)
 =\frac{1}{i\nu_n-\epsilon^{\rm d/h}_{j}},
 \label{eq:atomic-propagator}
\end{equation}
where $\omega_n$ and $\nu_n$ are the Matsubara
frequency for bosons and fermions, respectively.
Note that the atomic limit propagators of slave bosons are independent
of the spin. 
This means that the atomic limit exhibits spin-incoherent paramagnetism.
Hereafter we set $\hbar=1$.

Let us suppose the mean-field (MF) $\lambda_j$ to be determined by
the atomic limit.
Namely, the self-consistent equation for $\lambda_j$ corresponds
to the statistical average of the constraint~(\ref{eq:Constraint}) in
the atomic limit:
\begin{equation}
 2b(\epsilon_{{\rm b},j})+f(\epsilon_{{\rm h},j})+f(\epsilon_{{\rm d},j})=1, 
 \label{eq:SCE-lambda}
\end{equation}
where the prefactor $2$ comes from the spin degrees of freedom.
$f(\epsilon)$ and $b(\epsilon)$ are, respectively, the Fermi and Bose
distribution functions. 
One can expect that if the effect of the kinetic energy $H_{\rm K}$ is
small, that is, at relatively high temperature compared to the kinetic energy,
the validity of this treatment should be guaranteed.
We thus use the MF assumption for the Lagrange multiplier:
$\lambda_j\rightarrow\lambda$.
Simultaneously the local potentials are also replaced by
the homogeneous ones:
$\epsilon^{\rm x}_{j}\rightarrow \epsilon_{\rm x}$
where ${\rm x}={\rm b},{\rm h}, {\rm d}$.
Corresponding to the MF treatment for the Lagrange multiplier, the
atomic propagators also become independent of sites: For example, 
the replacement of $\lambda_j\rightarrow\lambda$ leads
the slave boson propagator 
$\mathcal{G}_{\mathrm{b}\sigma}^{(0)}(\bm{r}_j,i\omega_n)\rightarrow\bar{\mathcal{G}}_{\mathrm{b}}(i\omega_n)$
where the site-independent propagator is defined as
\begin{equation}
 \bar{\mathcal{G}}_{\mathrm{b}}(i\omega_n)
 =\frac{1}{i\omega_n-\epsilon_{\mathrm{b}}}.
 \label{eq:mean-field-spinon-propagator}
\end{equation}

\begin{figure*}[t]
 \begin{center}
  \includegraphics[scale=.8]{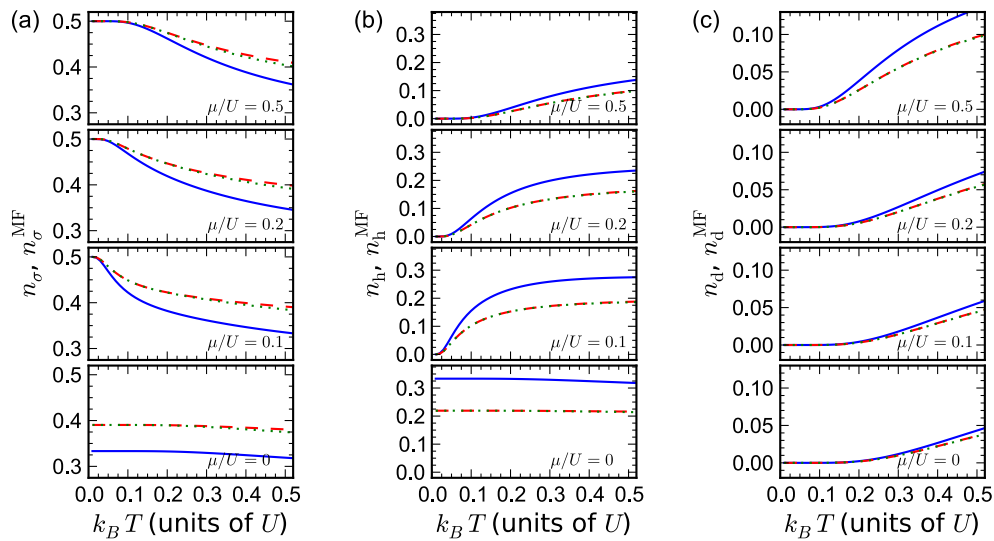}
  \caption{(cClor online) 
  The slave-particle densities as a function of temperature for
  different chemical potentials:
  (a) the slave boson, 
  (b) holon, and 
  (c) doublon density. 
  The chemical potentials are chosen as follows: $\mu/U=0.5$, $0.2$,
  $0.1$, and $0$ from the top to bottom.
  The MF results $n_{\sigma}^{\mathrm{MF}}$,
  $n_{\mathrm{h}}^{\mathrm{MF}}$, and $n_{\mathrm{d}}^{\mathrm{MF}}$
  given by the slave-particle technique and the exact results in the
  atomic limit, $n_{\sigma}$, $n_{\mathrm{h}}$, and $n_{\mathrm{d}}$, are
  compared. 
  $n_{\sigma}^{\mathrm{MF}}$, $n_{\mathrm{h}}^{\mathrm{MF}}$, and
  $n_{\mathrm{d}}^{\mathrm{MF}}$ are calculated numerically and
  analytically. 
  The solid, dashed, and dotted lines, respectively, denote the exact
  result by the atomic limit calculation, the slave particle
  approach~(\ref{eq:MF-slave}).}
  \label{fig:slave-density}
 \end{center}
\end{figure*}

Let us solve the self-consistent equation~(\ref{eq:SCE-lambda}).
For $k_{B}T/U\ll 1$, Eq.~(\ref{eq:SCE-lambda}) can be simplified, and
solved analytically as follows:
\begin{equation}
 \lambda
 =k_{B}T\log\frac{3+\sqrt{9+8(e^{-(U-\mu)/k_{B}T}+e^{-\mu/k_{B}T})}}{2}.
 \label{eq:MF-Lagrange-multiplier}
\end{equation}
As discussed below, Eq.~(\ref{eq:SCE-lambda}) is numerically solved, and
we compare the numerical result with
Eq.~(\ref{eq:MF-Lagrange-multiplier}). 
The obtained $\lambda$ leads the estimation of the slave particle
densities. 
The density of each slave particle is given by the MF solution
$\lambda$: 
$n_{\sigma}^{\mathrm{MF}}=b(\epsilon_{\mathrm{b}})$ for a slave boson, 
$n_{\mathrm{h}}^{\mathrm{MF}}=f(\epsilon_{\mathrm{h}})$ for a holon and 
$n_{\mathrm{d}}^{\mathrm{MF}}=f(\epsilon_{\mathrm{d}})$ for a doublon. 
The temperature and chemical potential dependency of them are shown in
Fig.~\ref{fig:slave-density}. 
In the temperature region shown in Fig.~\ref{fig:slave-density}, the
results analytically given by Eq.~(\ref{eq:MF-Lagrange-multiplier}) are
in precise agreement with ones given by the numerically solved
$\lambda$.
It means that in such a regime we may always employ
Eq.~(\ref{eq:MF-Lagrange-multiplier}) as a solution of the
MF self-consistent equation~(\ref{eq:SCE-lambda}). 
On the other hand, the densities in the atomic limit can be exactly
calculated as shown in Appendix~\ref{sec:app2}. 
Here we take the exact densities to be $n_{\sigma}$, $n_{\mathrm{h}}$
and $n_{\mathrm{d}}$, and their temperature and chemical potential
dependency is also shown in Fig.~\ref{fig:slave-density}.
The exact result allows us to discuss the temperature and chemical
potential regime justifying the slave particle approach. 
From the comparison made in Fig.~\ref{fig:slave-density}, the slave
particle technique is found to agree with the exact result in the
temperature and chemical potential region where
$n_{\sigma}^{\mathrm{MF}}=1/2$ and
$n_{\mathrm{h}}^{\mathrm{MF}}=n_{\mathrm{d}}^{\mathrm{MF}}=0$.
In other words the slave particle technique is expected to be reasonable
when the system is near a Mott insulator, and such  temperature as a
benchmark is below $k_{B}T/U\approx 0.1$.

% \begin{figure}[htbp]
%  \begin{center}
%   \includegraphics[scale=0.7]{density.eps}
%   \caption{(color online). The chemical potential dependence of the
%   density for a doublon (dotted), holon (dashed-dotted), slave boson
%   (dashed) and original fermion (solid).
%   The thin and thick lines denote $k_{\rm B}T/U=0.05$ and $0.1$,
%   respectively}
%   \label{fig:chemical_potential_dependence}
%  \end{center}
% \end{figure}

\subsection{The finite hopping}
\begin{figure}[b]
 \begin{center}
  \includegraphics[scale=0.9]{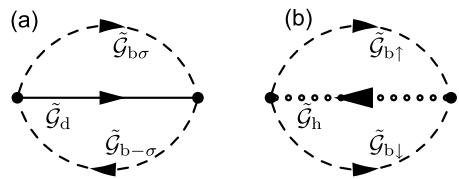}
  \caption{The NC diagrams giving the doublon self-energy.
  The solid, double dotted, and dashed lines, respectively, denotes the
  full propagators of the doublon, holon, and slave bosons.
  The left diagram (a) describes the scattering between a doublon and
  slave boson.
  The right diagram (b) represents the higher energy scattering to a
  holon than the left (a).
  Thus, as long as charge excitations of energy $\sim U$ are taken
  for large $U$, the diagram (b) would be irrelevant.
  The holon self-energy is also given by the same type of diagrams.}
\label{fig:non-crossing-diagram}
 \end{center}
\end{figure}

We now consider a finite but small hopping by taking the infinite series
of diagrams produced by the scattering $H_{\rm K}$ among the slave
particles, based on the non-crossing (NC)
approximation~\cite{Brinkman.Rice/PRB2.1970,Kane.Lee.Read/PRB39.1989}.
This approximation can be also regarded as a certain type of high
temperature series expansion (HTSE)~\cite{Henderson.etal/PRB.46.1992},
but in our formalism the Wick theorem is still applicable due to the
quadratic Hamiltonian~(\ref{eq:AtomicHamiltonian}), and a particular
infinite series of kinetic energy perturbation can be taken.
Based on the NC approximation the self-energy diagrams constructed by
full propagators, shown in Fig.~\ref{fig:non-crossing-diagram}, are
considered. 

\begin{figure}[t]
 \begin{center}
  \includegraphics[scale=0.7]{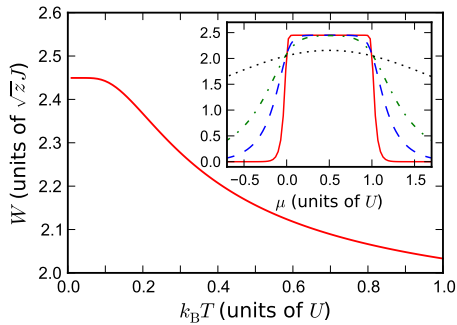}
  \caption{(Color online) The doublon-holon band width $W/\sqrt{z}J$ as
  a function of $k_{\rm B}T$ for $\mu/U=0.5$.
  The chemical potential dependence of the band width for
  $k_{\rm B}T/U=0.01$ (solid), $0.05$ (dashed), $0.1$ (dashed-dotted),
  and $0.5$ (dotted) is also shown in the inset.}
  \label{fig:band-width}
 \end{center}
\end{figure}

Since one is in a temperature regime higher than the anti-ferromagnetic
exchange $\sim 4J^2/U$,
we apply the spin-incoherent assumption to the slave boson propagator.
Namely, the slave boson propagators in the diagrams of
Fig.~\ref{fig:non-crossing-diagram} are replaced by the atomic
one~(\ref{eq:mean-field-spinon-propagator}),
\begin{equation}
 \tilde{\mathcal{G}}_{\rm{b}\sigma}(\bm{k},i\omega_n)
  \rightarrow
  \bar{\mathcal{G}}_{\mathrm{b}}(i\omega_n).
  \label{eq:spin-incoherent-assumption}
\end{equation}
The doublon and holon propagators are left full ones:
namely, by using the Dyson equation the full propagators are written as 
\begin{align}
 \tilde{\mathcal{G}}_{\mathrm{d/h}}(\bm{k},i\nu_n)
 &=\tilde{\mathcal{G}}^{(0)}_{\mathrm{d/h}}(\bm{k},i\nu_n)
 \nonumber \\
 &\quad 
 +\tilde{\mathcal{G}}^{(0)}_{\mathrm{d/h}}(\bm{k},i\nu_n)
  \tilde{\Sigma}_{\mathrm{d/h}}(\bm{k},i\nu_n)
  \tilde{\mathcal{G}}_{\mathrm{d/h}}(\bm{k},i\nu_n).
 \label{eq:dyson-equation}
\end{align}
Unlike the standard MF theory, the dynamical fluctuation of the
slave bosons is retained here, which is necessary to describe the
doublon-holon excitation.
The NC diagram Fig.~\ref{fig:non-crossing-diagram} (b)
couples the self-consistent equations of the doublon and holon self-energy.
However, the contribution is negligibly small because it is a far
off-shell diagram in this case such as Mott state.
Consequently the self-consistent equations of the self-energy
Fig.~\ref{fig:non-crossing-diagram}~(a) are decoupled and one can 
obtain in momentum space,
\begin{equation}
 \tilde{\Sigma}_{\rm d}(\bm{k},i\nu_n)
 =\frac{W^2}{4}
   \frac{1}{N}
   \sum_{\bm{p}}\tilde{\mathcal{G}}_{\rm d}(\bm{p},i\nu_n).
   \label{eq:SCE-selfenergy}
\end{equation}
with $W=\sqrt{8zb(\epsilon_{\rm b})[b(\epsilon_{\rm b})+1]J^2}$
corresponding to a half band width for the holon and doublon as we will 
see below.
$z$ is a coordination number, and $N$ is the total site number of the
system.
Note that due to the momentum dependence of the right-hand side the
self-energy should be given as a local quantity:
$\tilde{\Sigma}_{\rm d}(\bm{k},i\nu_n)=\Sigma_{\rm d}(i\nu_n)$.
Consequently the propagator also turns out to be local:
$\tilde{\mathcal{G}}_{\mathrm{d}}(\bm{k},i\nu_n)=\mathcal{G}_{\mathrm{d}}(i\nu_n)$.
Thus the self-consistent equation~(\ref{eq:SCE-selfenergy}) is easily
solved through the Dyson equation~(\ref{eq:dyson-equation}) as follows:  
\begin{equation}
 \Sigma_{\rm d}(i\nu_n)
 =\frac{
   i\nu_n-\epsilon_{\rm d}
   -i\sqrt{W^2-(i\nu_n-\epsilon_{\rm d})^2}
        }{2}.
\label{eq:selfenergy}
\end{equation}
Through the analytic continuation the doublon spectral function, which
is equivalent to the density of state (DOS) in this case, is also
obtained.
In this approximation, a semi-circle type DOS is formed:
\begin{equation}
 {\cal A}_{\rm d}(\omega)
 =\frac{4}{W}
  \sqrt{1-\left(\frac{\omega-\epsilon_{\rm d}}{W}\right)^2}.
\label{eq:doublon-spectralfunction}
\end{equation}
One can also obtain the self-energy and spectral function of the holon
in the same way. 
The forms are the same as what is obtained by the replacement
$\epsilon_{\mathrm{d}}\rightarrow\epsilon_{\mathrm{h}}$ in 
Eqs.~(\ref{eq:selfenergy}) and (\ref{eq:doublon-spectralfunction}).
The holon spectral function is found to reasonably reproduce the result
of Brinkman and Rice on single hole dynamics in a Mott
insulator~\cite{Brinkman.Rice/PRB2.1970}.
The chemical potential dependency of the band width $W$ is
shown in Fig.~\ref{fig:band-width}.

\begin{figure}[bp]
 \begin{center}
  \includegraphics[scale=0.7]{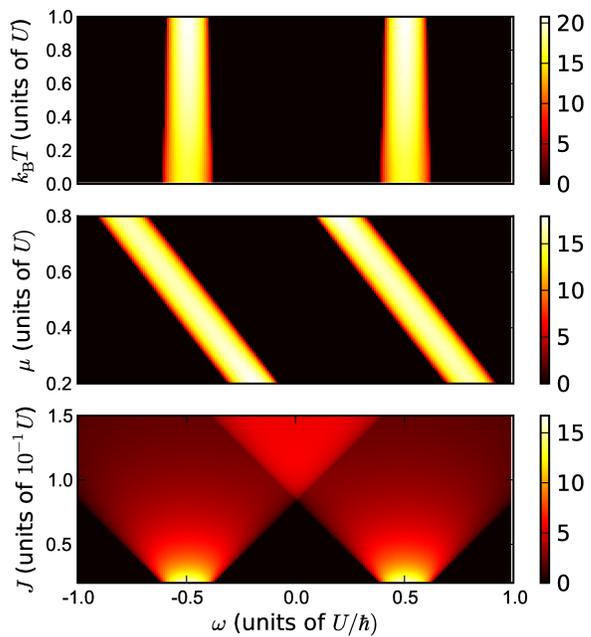}
  \caption{(Color online) The spectral function of the original fermion
  as a function of $k_{\rm B}T/U$ for $\mu/U=0.5$ and $J/U=0.02$ 
  (top panel), $\mu/U$ for $k_{\rm B}T/U=0.1$, and $J/U=0.02$ 
  (middle panel) and $J/U$ for $k_{\rm B}T/U=0.1$ and $\mu/U=0.5$
  (bottom panel).} 
  \label{fig:spec.func.}
 \end{center}
\end{figure}

Using the representation (\ref{eq:SF-SB}), via the Matsubara Green's
function of the original fermion, the DOS (spectral function) is
represented as
\begin{align}
 {\cal A}_{\sigma}(\omega)
 &=\biggl[b(\epsilon_{\rm b})+f(\epsilon_{\rm b}-\omega)\biggr]
    {\cal A}_{\rm h}(\epsilon_{\rm b}-\omega)
 \nonumber \\
 &\quad
   +\biggl[b(\epsilon_{\rm b})+f(\epsilon_{\rm b}+\omega)\biggr]
     {\cal A}_{\rm d}(\epsilon_{\rm b}+\omega).
\end{align}
As expected, the doublon and holon spectral functions,
${\cal A}_{\rm d}$ and ${\cal A}_{\rm h}$, give the upper and
lower Hubbard band, respectively.
The spectral function as a function of $J/U$ and the chemical potential
is shown in Fig.\ref{fig:spec.func.}.

\section{Doublon production rate}\label{sec:DPR}
\begin{figure*}[t]
 \begin{center}
  \includegraphics[scale=.8]{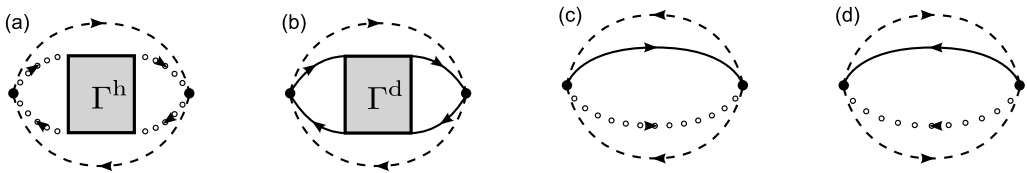}
  \caption{The diagrams contributing to the considered correlation
  function~(\ref{eq:H_K-correlation-function0}): The diagrams (a), (b),
  (c) and (d) correspond to the first, second, third, and fourth terms. 
  The solid, double dotted, and dashed lines denotes the doublon, holon,
  and slave boson propagators, respectively.}
  \label{fig:kinetic-correlation-function}
 \end{center}
\end{figure*}
We calculate the DPR induced by the amplitude modulation of an
optical lattice potential using the above formalism.
The amplitude modulation of the optical lattice potential,
$V(t)=V_0+\delta{V}\cos(\omega t)$, modifies both $J$ and $U$ as
$J\rightarrow J[1+\delta{J}\cos(\omega t)]$ and
$U\rightarrow U[1+\delta{U}\cos(\omega t)]$, where $\delta{J}$ and
$\delta{U}$ are dimensionless modulation perturbation parameters and
given as a function of $\delta{V}$.
However, it is possible to map the two parameter
modulations problem to single parameter one of either $J$ or
$U$~\cite{Reischl.etal/PRA72.2005,Iucci.etal/PRA73R.2006,Kollath.etal/RPA74R.2006,Sensarma.etal/PRL103.2009}.
Namely, the modulation perturbation to be discussed here can be written
as
\begin{equation}
 H_{\rm mod}(t)=\delta{F}\cos(\omega t)H_{\rm K},
\end{equation} 
where $\delta{F}=\delta{J}-\delta{U}$.~\footnote{There is a slight additional modification coming from the presence of the trap in this
formula, but this does not change quantitatively the results for the case considered here}
Within the second-order perturbation in terms of $\delta{F}$, the
DPR defined as the time averaged growth rate of
{\it atoms forming doublons} is given as~\cite{Kollath.etal/RPA74R.2006}
\begin{equation}
 P_{\rm D}(\omega)
 =-\frac{(\delta{F})^2}{U}
   \omega\Im{{\chi}^{\rm R}_{\rm K}(\omega)},
\end{equation}
where
${\chi}^{\rm R}_{\rm K}(\omega)=-i\int_0^\infty\!\!dt\ e^{i\omega t}\langle[H_{\rm K}(t),H_{\rm K}(0)]\rangle$.
As shown in Appendix~\ref{sec:app1},
we represent the correlation function
$\chi_{\mathrm{K}}(\tau)=-\langle{T_{\tau}H_{\mathrm{K}}(\tau)H_{\mathrm{K}}(0)}\rangle$
without vertex corrections in the slave-particle description~\footnote{The vertex corrections give at least the
correction of ${\cal O}((J/U)^4)$. In addition, it does not affect
crucially in the insulator with the large Mott gap.}, 
\begin{align}
 \chi_{\rm K}(\tau)
 &=-2J^2
   \sum_{\langle{i,j}\rangle}
   \{
     \Gamma_{ij}^{\mathrm{h}}(\tau)\bar{\mathcal{G}}_{\mathrm{b}}(\tau)\bar{\mathcal{G}}_{\mathrm{b}}(-\tau)
     +\Gamma_{ij}^{\mathrm{d}}(\tau)\bar{\mathcal{G}}_{\mathrm{b}}(\tau)\bar{\mathcal{G}}_{\mathrm{b}}(-\tau)
 \nonumber \\
 & \quad
     -[\bar{\mathcal{G}}_{\mathrm{b}}(-\tau)]^2
      \mathcal{G}_{\mathrm{h}}(\tau)\mathcal{G}_{\mathrm{d}}(\tau)
     -[\bar{\mathcal{G}}_{\mathrm{b}}(\tau)]^2
      \mathcal{G}_{\mathrm{h}}(-\tau)\mathcal{G}_{\mathrm{d}}(-\tau)
   \},
  \label{eq:H_K-correlation-function0}
\end{align}
% \begin{widetext}
% \begin{equation}
%  \chi_{\rm K}(\tau)
%  =-2J^2
%    \sum_{\langle{i,j}\rangle}
%    \biggl[
%      \Gamma_{ij}^{\mathrm{h}}(\tau)\bar{\mathcal{G}}_{\mathrm{b}}(\tau)\bar{\mathcal{G}}_{\mathrm{b}}(-\tau)
%      +\Gamma_{ij}^{\mathrm{d}}(\tau)\bar{\mathcal{G}}_{\mathrm{b}}(\tau)\bar{\mathcal{G}}_{\mathrm{b}}(-\tau)
%      -\Bigl(\bar{\mathcal{G}}_{\mathrm{b}}(-\tau)\Bigr)^2
%       \mathcal{G}_{\mathrm{h}}(\tau)\mathcal{G}_{\mathrm{d}}(\tau)
%      -\Bigl(\bar{\mathcal{G}}_{\mathrm{b}}(\tau)\Bigr)^2
%       \mathcal{G}_{\mathrm{h}}(-\tau)\mathcal{G}_{\mathrm{d}}(-\tau)
%    \biggr],
%   \label{eq:H_K-correlation-function0}
% \end{equation}
where $\Gamma_{ij}^{\mathrm{X}}(\tau)=\langle{T_{\tau}x_{i}^\dagger(\tau)x_j(\tau)x_j^\dagger(0)x_i(0)}\rangle$
($x=d$ or $h$) is a two-particle Green's function of a doublon
($\mathrm{X}=\mathrm{d}$) and holon ($\mathrm{X}=\mathrm{h}$). 
The diagrams corresponding to the terms in
Eq.~(\ref{eq:H_K-correlation-function0}) are illustrated in
Fig.~\ref{fig:kinetic-correlation-function}.
Without the vertex correction,
the two-particle propagators are contracted to a single particle
propagators by the Wick expansion:
$\Gamma_{ij}^{\rm X}(\tau)=-{\cal G}_{\rm X}(\tau){\cal G}_{\rm X}(-\tau)$.
Through the Fourier transform of $\chi_{\rm K}(\tau)$ and analytic
continuation, one can straightforwardly obtain the real-time kinetic
energy correlation function in frequency domain.
As a result, the imaginary part of the correlation function is given as
\begin{align}
 \frac{\Im{\chi_{\rm K}^{\rm R}(\omega)}}{-NW^2/8}
 &=\int\!\!\frac{d\nu}{2\pi}
      [f(\nu-\omega)-f(\nu)]
      [{\cal A}_{\rm h}(\nu){\cal A}_{\rm h}(\nu-\omega)
 \nonumber \\
 & \quad
        +{\cal A}_{\rm d}(\nu){\cal A}_{\rm d}(\nu-\omega)]
        +2\sinh(\epsilon_{\rm b})
          \int\!\!\frac{d\nu}{2\pi}
              [b(2\epsilon_{\rm b})
 \nonumber \\
 & \quad
               +f(\nu)]
    \{
     [f(2\epsilon_{\rm b}-\nu)-f(2\epsilon_{\rm b}-\nu+\omega)]
 \nonumber \\
 & \quad
     \times{\cal A}_{\rm d}(\nu){\cal A}_{\rm h}(2\epsilon_{\rm b}+\omega-\nu)
      -(\omega\rightarrow -\omega)
    \}.
\label{eq:H_K-correlation-function}
\end{align}
% \begin{align}
%  \frac{\Im{\chi_{\rm K}^{\rm R}(\omega)}}{-NW^2/8}
%  &=\int\!\!\frac{d\nu}{2\pi}
%       \biggl(
%         f(\nu-\omega)-f(\nu)
%       \biggr)
%       \biggl(
%         {\cal A}_{\rm h}(\nu){\cal A}_{\rm h}(\nu-\omega)
%         +{\cal A}_{\rm d}(\nu){\cal A}_{\rm d}(\nu-\omega)
%       \biggr)
%  \nonumber \\
%  &\quad
%   +2\sinh(\epsilon_{\rm b})
%    \int\!\!\frac{d\nu}{2\pi}
%      \biggl[
%        b(2\epsilon_{\rm b})+f(\nu)
%      \biggr]
%     \biggl[
%      \biggl\{
%        f(2\epsilon_{\rm b}-\nu)-f(2\epsilon_{\rm b}-\nu+\omega)
%      \biggr\}
%      {\cal A}_{\rm d}(\nu)
%      {\cal A}_{\rm h}(2\epsilon_{\rm b}+\omega-\nu)
%       -\biggl(\omega\rightarrow -\omega \biggr)
%     \biggr].
% \label{eq:H_K-correlation-function}
% \end{align}
% \end{widetext}

\section{Comparison with experiments}\label{sec:comparison-experiments}
Let us compare our result~(\ref{eq:H_K-correlation-function}) with
the experimental data.
We employ a set of the parameters evaluated in the $^{40}$K atom
experiment~\cite{Greif.etal/PRL.106.2011}:
the hopping $J/\hbar=2\pi\times 85$ (Hz) and
the interaction $U/\hbar=2\pi\times 5400$ (Hz).
In terms of the optical lattice potential, the depth, modulation
rate and lattice constant are, respectively, taken to be
$V_0=10E_{\rm R}$ where $E_{\rm R}$ is a recoil energy,
$\delta{V}/V_0=0.1$ and $a=532$ (nm).
The lattice modulation is translated into $\delta{F}\approx -0.32$ in
hopping modulation.
The LDA is used to take into account the
effect of the harmonic trap potential $V_{\rm trap}(\bm{r})$ whose
frequency is
$(\omega_x/2\pi,\omega_y/2\pi,\omega_z/2\pi)=(56,61,139)$ (Hz).
In the LDA, we replace the chemical potential of the homogeneous case
by the local one, $\mu(\bm{r})=\mu_0-V_{\rm trap}(\bm{r})$ where
$\mu_0$ is self-consistently determined to give the total trapped atom
number $8\times 10^{4}$.
In our framework, temperature is treated as a free parameter so that
we determine the temperature by a fit of the DPR spectrum intensity at
$\omega=U/\hbar$, which is obtained in the experiment.
\begin{figure}[tbp]
 \begin{center}
  \includegraphics[scale=0.7]{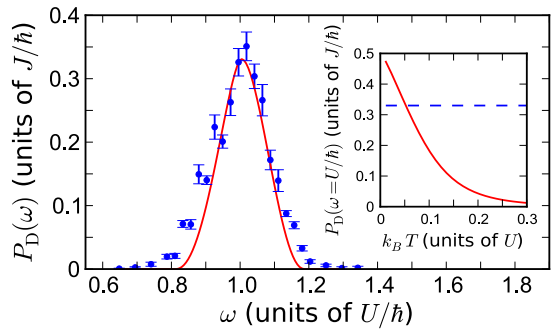}
  \caption{(Color online) The DPR spectrum as a function of modulation
  frequency.  
  The solid line and points denote the theoretical and experimental
  results, respectively.
  The temperature necessary to draw the theoretical curve is determined
  in the inset; the temperature dependence of the theoretically given
  DPR at $\omega=U/\hbar$ and the temperature is determined from the
  crossing point to the experimental data (dotted line), which is
  $k_{\rm B}T/U\approx0.052$.
  }
  \label{fig:DPR}
 \end{center}
\end{figure}

The temperature dependence of the DPR spectrum at $\omega=U/\hbar$ is
shown in the inset in Fig.~\ref{fig:DPR}, and 
$k_{\rm B}T/U\approx 0.052$ in this system is determined.\footnote{A
least square fit has been also implemented, and gives the similar
result $k_{B}T/U\approx 0.045$.} 
The determined temperature is in the region to justify the
slave-particle approach, which is discussed in
Sec.~\ref{sec:diagrammatic-analys}, and the theory is thus expected to
work well.
Furthermore, using the obtained temperature, we plot the DPR spectrum in
Fig.~\ref{fig:DPR}. 
The agreement is remarkably good. 
In addition to giving access to the line shape it means that via our
theory one can use the shaking method as a good thermometer in low
temperature regime~\footnote{In contrast the HTSE would be instead
better in the higher temperature region, as discussed in
Ref.~\cite{Joerdens.etal/PRL.104.2010}}, since the 
curve giving the amplitude versus temperature (the inset in
Fig.~\ref{fig:DPR}) is reasonably smooth and steep. To check this point
we compare in our case the temperature determined by the fitting of the
shaking curve with other estimates from entropy
calculations~\cite{Joerdens.etal/PRL.104.2010} and find that the two
results are perfectly consistent.

\section{Conclusion and summary}\label{sec:summary}
In this paper, we have described the charge excitation of strongly
correlated fermionic systems in the spin-incoherent paramagnetic regime
by a slave-particle representation and 
diagrammatic approach from the atomic limit. 
This method allows us to take the finite temperature and trapping into
account. 
Based on the spectrum functions of the doublon and holon, the analytical
form of the DPR spectrum as a second-order response of the optical
lattice modulation has been given, and extended
to the homogeneous system of the trapped atom cloud by using the local
density approximation. 
In addition, substituting the parameters evaluated in the
experiment~\cite{Greif.etal/PRL.106.2011}, a comparison with the
experiment has been made.
Although temperature has not been directly measured in experiments, it
has been determined as an optimization parameter which is controlled to
fit the experimental data. 
The result has been in  agreement with the experimental data, which
shows that one can use the lattice modulation spectroscopy as a
thermometer.

Our method has potential extension such as
SU($N$) higher symmetric atom systems realized in alkaline-earth-metal atom
experiments.~\cite{Cazalilla.Ho.Ueda/New.J.Phys.11.2009,Taie.etal/PRL.105.2010,Gorshkov.etal/NatPhys.6.2010,DeSalvo.etal/PRL.105.2010}

\acknowledgments
We are grateful to C. Berthod, T. Esslinger, D. Greif, N. Kawakami,
A. Koga, L. Pollet, L. Tarruell, and T. Uehlinger for fruitful
discussions. 
In addition, we thank D. Pekker for contributions at an early stage of this work
and T. Esslinger's group for making their data available.
ED acknowledges support from the DARPA OLE program, Harvard-MIT CUA, NSF
Grant No. DMR-07-05472, AFOSR Quantum Simulation MURI. 
This work was supported by the Swiss National Foundation under MaNEP and
division II.

\appendix
 \section{The slave-particle density in the atomic limit}
 \label{sec:app2} 
 In order to clarify the parameter regime to justify the slave-particle
 approach, we make a comparison with the exact calculation in the atomic
 limit. 
 Then we focus on the single site occupations as a
 function of temperature and chemical potential, which is identical to
 the densities of the slave boson, holon, and doublon. 
 
 Let us perform the atomic limit calculation. 
 We consider the atomic Hamiltonian~(\ref{eq:atomic-Hamiltonian}).
 Then the partition function is calculated as 
 \begin{equation}
  Z_{\mathrm{atom}}
   =1+2x+x^2y
   \label{eq:atomic-limit-partition-function}
 \end{equation}
 where $x=e^{\mu/K_{B}T}$ and $y=e^{-U/k_{B}T}$.
 To calculate the densities of the slave boson, holon, and doublon, 
 we introduce the projection operators. 
 In the original fermion picture, the projection operators of the slave boson,
 holon and doublon state are, respectively, written as 
 $p_{\sigma}=n_{\sigma}-n_{\uparrow}n_{\downarrow}$,
 $p_{\mathrm{h}}=(n_{\uparrow}-1)(n_{\downarrow}-1)$,
 and $p_{\mathrm{d}}=n_{\uparrow}n_{\downarrow}$.
 They obviously obeys the constraint
 $\sum_{\sigma}p_{\sigma}+p_{\mathrm{h}}+p_{\mathrm{d}}=1$.
 Taking the statistical average of these projection operators, we obtain
 the following exact result of slave-particle densities: 
 \begin{subequations}
  \begin{align}
   n_{\sigma}&=\langle{p_{\sigma}}\rangle=\frac{x}{1+2x+x^2y}, \\
   n_{\mathrm{h}}&=\langle{p_{\mathrm{h}}}\rangle=\frac{1}{1+2x+x^2y}, \\
   n_{\mathrm{d}}&=\langle{p_{\mathrm{d}}}\rangle=\frac{x^2y}{1+2x+x^2y}.
  \end{align}
  \label{eq:atom-limit-calc}
 \end{subequations}

 On the other hand, in the case of the slave-particle approach combined
 with the MF solution of the Lagrange multiplier $\lambda$, the
 slave-particle densities are given as
 \begin{subequations}
  \begin{align}
   n_{\sigma}^{\mathrm{MF}}=b(\epsilon_{\mathrm{b}}), \\
   n_{\mathrm{h}}^{\mathrm{MF}}=f(\epsilon_{\mathrm{h}}), \\
   n_{\mathrm{d}}^{\mathrm{MF}}=f(\epsilon_{\mathrm{d}}),
  \end{align}
  \label{eq:MF-slave}
 \end{subequations}
 where 
 $\epsilon_{\mathrm{b}}=\lambda$,
 $\epsilon_{\mathrm{h}}=\mu+\lambda$, and
 $\epsilon_{\mathrm{d}}=U-\mu+\lambda$, and $\lambda$ is a solution of
 the MF constraint equation~(\ref{eq:SCE-lambda}).
 For $k_{B}T/U\ll 1$ the MF Lagrange
 multiplier $\lambda$ is analytically obtained as shown in
 Eq.~(\ref{eq:MF-Lagrange-multiplier}).
 The comparison of the exact result~(\ref{eq:atom-limit-calc}) and the
 MF approach~(\ref{eq:MF-slave}) is shown in
 Fig.~\ref{fig:slave-density}.

 \section{Calculation of the kinetic-energy correlation function}
 \label{sec:app1}
 Here we show the calculation of the kinetic energy correlation function
 $\chi_{\mathrm{K}}(\tau)$ in detail, and derive
 Eq.~(\ref{eq:H_K-correlation-function0}) given in Sec.~\ref{sec:DPR}. 
 Using the expression of the kinetic energy
 Hamiltonian~(\ref{eq:TunnelingHamiltonian}), 
 the correlation function $\chi_{\mathrm{K}}(\tau)$ is straightforwardly
 written as 
  \begin{align}
   \chi_{\mathrm{K}}(\tau)
   &=-J^2
     \sum_{\langle{i,j}\rangle,\langle{l,m}\rangle}
     \{
       \langle
         T_{\tau}F_{ji}(\tau)F_{ml}(0)
       \rangle
       \langle
         T_{\tau}
         [h_{i}^{\dagger}(\tau)h_{j}(\tau)
   \nonumber \\
   & \quad
          -d_{i}^{\dagger}(\tau)d_{j}(\tau)]
         [h_{l}^\dagger(0)h_{m}(0)-d_{l}^{\dagger}(0)d_{m}(0)]
        \rangle
   \nonumber \\
   & \quad
   +
   \langle
     T_{\tau}A_{ij}^{\dagger}(\tau)A_{lm}(0)
   \rangle
   \langle
     T_{\tau}d_{j}(\tau)d_{m}^{\dagger}(0)
   \rangle
   \langle
     T_{\tau}h_{i}(\tau)h_{l}^{\dagger}(0)
   \rangle
   \nonumber \\
   & \quad
   +
   \langle
     T_{\tau}A_{ij}(\tau)A_{lm}^{\dagger}(0)
   \rangle
   \langle
     T_{\tau}h_{i}^{\dagger}(\tau)h_{m}(0)
   \rangle
   \langle
     T_{\tau}d_{j}^{\dagger}(\tau)d_{l}(0)
   \rangle
   \},
  \label{eq:kinetic-energy-correlation}
  \end{align} 
%% \begin{widetext}
  % \begin{align}
  %  \chi_{\mathrm{K}}(\tau)
  %  &=-J^2
  %    \sum_{\langle{i,j}\rangle,\langle{l,m}\rangle}
  %    \biggl[
  %      \langle
  %        T_{\tau}F_{ji}(\tau)F_{ml}(0)
  %      \rangle
  %      \langle
  %        T_{\tau}
  %        [h_{i}^{\dagger}(\tau)h_{j}(\tau)-d_{i}^{\dagger}(\tau)d_{j}(\tau)]
  %        [h_{l}^\dagger(0)h_{m}(0)-d_{l}^{\dagger}(0)d_{m}(0)]
  %       \rangle
  %  \nonumber \\
  %  & \quad
  %  +
  %  \langle
  %    T_{\tau}A_{ij}^{\dagger}(\tau)A_{lm}(0)
  %  \rangle
  %  \langle
  %    T_{\tau}d_{j}(\tau)d_{m}^{\dagger}(0)
  %  \rangle
  %  \langle
  %    T_{\tau}h_{i}(\tau)h_{l}^{\dagger}(0)
  %  \rangle
  %  +
  %  \langle
  %    T_{\tau}A_{ij}(\tau)A_{lm}^{\dagger}(0)
  %  \rangle
  %  \langle
  %    T_{\tau}h_{i}^{\dagger}(\tau)h_{m}(0)
  %  \rangle
  %  \langle
  %    T_{\tau}d_{j}^{\dagger}(\tau)d_{l}(0)
  %  \rangle
  %  \biggl],
  % \label{eq:kinetic-energy-correlation}
  % \end{align} 
 where the correlation functions between the
 operators $F_{ji}$ and $A_{ml}$ have been supposed to be zero because
  the slave bosons do not condense in this case.
 The autocorrelations for $F_{ji}$ and $A_{ml}$ should be finite, and in
 the calculation the spin-incoherent
 assumption~(\ref{eq:spin-incoherent-assumption}) is applied in the 
 same way as in Sec.~\ref{sec:diagrammatic-analys}. 
 As a result, the autocorrelations are calculated as follows: 
\begin{align}
 &\langle{T_{\tau}F_{ji}(\tau)F_{lm}(0)}\rangle
 =2\delta_{j,l}\delta_{i,j}
   \bar{\mathcal{G}}_{\mathrm{b}}(-\tau)
   \bar{\mathcal{G}}_{\mathrm{b}}(\tau),
 \label{eq:ff-correlation}\\
 &\langle{T_{\tau}A_{ij}(\tau)A_{lm}^{\dagger}(0)}\rangle
 =-2\left(\delta_{j,l}\delta_{i,m}+\delta_{j,m}\delta_{i,l}\right)
   \bar{\mathcal{G}}_{\mathrm{b}}(\tau)\bar{\mathcal{G}}_{\mathrm{b}}(\tau).
 \label{eq:aa-correlation2}
\end{align}
As seen in Sec.~\ref{sec:diagrammatic-analys} the
doublon and holon propagators turn out to be local: Namely,
\begin{align}
 &\langle{T_{\tau}h_{i}(\tau)h_{j}^{\dagger}(0)}\rangle
 =i\delta_{i,j}\mathcal{G}_{\mathrm{h}}(\tau),
 \label{eq:holon-propagator} \\
 &\langle{T_{\tau}d_{i}(\tau)d_{j}^{\dagger}(0)}\rangle
 =i\delta_{i,j}\mathcal{G}_{\mathrm{d}}(\tau).
 \label{eq:doublon-propagator}
\end{align}
 We substitute
 Eqs.~(\ref{eq:ff-correlation})-~(\ref{eq:doublon-propagator}) into
 Eq.~(\ref{eq:kinetic-energy-correlation}), and then the correlation
  function is rewritten as
 \begin{align}
  \chi_{\mathrm{K}}(\tau)
  &
  =-2J^2\sum_{\langle{i,j}\rangle}
   \{
     \bar{\mathcal{G}}_{\mathrm{b}}(-\tau)
     \bar{\mathcal{G}}_{\mathrm{b}}(\tau)
     \left[
       \Gamma^{\mathrm{h}}_{i,j}(\tau)+\Gamma^{\mathrm{d}}_{i,j}(\tau)
     \right]
  \nonumber \\
  & \quad
     -
     [\bar{\mathcal{G}}_{\mathrm{b}}(-\tau)]^2
     \mathcal{G}_{\mathrm{h}}(\tau)
     \mathcal{G}_{\mathrm{d}}(\tau)
     -
     [\bar{\mathcal{G}}_{\mathrm{b}}(\tau)]^2
     \mathcal{G}_{\mathrm{h}}(-\tau)
     \mathcal{G}_{\mathrm{d}}(-\tau)
   \},
 \end{align}
 % \begin{align}
 %  \chi_{\mathrm{K}}(\tau)
 %  &
 %  =-2J^2\sum_{\langle{i,j}\rangle}
 %   \biggl[
 %     \bar{\mathcal{G}}_{\mathrm{b}}(-\tau)
 %     \bar{\mathcal{G}}_{\mathrm{b}}(\tau)
 %     \biggl(
 %       \Gamma^{\mathrm{h}}_{i,j}(\tau)+\Gamma^{\mathrm{d}}_{i,j}(\tau)
 %     \biggr)
 %     -
 %     \Bigl(\bar{\mathcal{G}}_{\mathrm{b}}(-\tau)\Bigr)^2
 %     \mathcal{G}_{\mathrm{h}}(\tau)
 %     \mathcal{G}_{\mathrm{d}}(\tau)
 %     -
 %     \Bigl(\bar{\mathcal{G}}_{\mathrm{b}}(\tau)\Bigr)^2
 %     \mathcal{G}_{\mathrm{h}}(-\tau)
 %     \mathcal{G}_{\mathrm{d}}(-\tau)
 %   \biggr],
 % \end{align}
%% \end{widetext}
 where the following two particle correlation functions have been defined as 
\begin{align}
 &\Gamma^{\mathrm{h}}_{i,j}(\tau)
 =\langle{T_{\tau}h_{i}^{\dagger}(\tau)h_{j}(\tau)h_{j}^{\dagger}(0)h_{i}(0)}\rangle,
 \\
 &\Gamma^{\mathrm{d}}_{i,j}(\tau)
 =\langle{T_{\tau}d_{i}^{\dagger}(\tau)d_{j}(\tau)d_{j}^{\dagger}(0)d_{i}(0)}\rangle.
\end{align}

\bibliographystyle{apsrev4-1}
% \bibliography{references}
%

\end{document}